\documentclass[twocolumn]{el-author}

\newcommand{\ua}{\uparrow}
\newcommand{\nc}{\newcommand}
\nc{\da}{\downarrow} \nc{\hc}{\hat{c}} \nc{\hS}{\hat{S}}
\nc{\bra}{\langle} \nc{\ket}{\rangle} \nc{\eq}{equation (\ref}
\nc{\h}{\hat} \nc{\hT}{\h{T}}\nc{\be}{\begin{eqnarray}}
	\nc{\ee}{\end{eqnarray}}\nc{\rd}{\textrm{d}}\nc{\e}{eqnarray}\nc{\hR}{\hat{R}}\nc{\Tr}{\mathrm{Tr}}
\nc{\tS}{\tilde{S}}\nc{\tr}{\mathrm{tr}}\nc{\8}{\infty}\nc{\lgs}{\bra\ua,\phi|}\nc{\rgs}{|\ua,\phi\ket}
\nc{\hU}{\hat{U}}\nc{\lfs}{\bra\phi|}\nc{\rfs}{|\phi\ket}\nc{\hZ}{\hat{Z}}\nc{\hd}{\hat{d}}\nc{\mD}{\mathcal{D}}
\nc{\bd}{\bar{d}}\nc{\bc}{\bar{c}}\nc{\mc}{\mathcal}\nc{\ea}{eqnarray}\nc{\mG}{\mathcal{G}}\nc{\bce}{\begin{center}}
	\nc{\ece}{\end{center}}

\usepackage{color,tikz,cancel}

\begin{document}

\title{The DoF Region of Two-User MIMO Broadcast Channel with Delayed Imperfect-Quality CSIT}

\author{Tong Zhang}

\abstract{The channel state information at the transmitter (CSIT) play an important role in the performance of wireless networks. The CSIT model can be delayed and imperfect-quality, since the feedback link has a delay and the channel state information (CSI) feedback has distortion. In this paper, we thus characterize the degrees-of-freedom (DoF) region of the two-user multiple-input multiple-output (MIMO) broadcast channel with delayed imperfect-quality CSIT, where the antenna configurations can be arbitrary. The converse proof of DoF region is based on the enhancement of physically degraded channel. The achievability proof of DoF region is through a novel transmission scheme design, where the duration of each phase and the amount of transmitted symbols are configured based on the imperfect-quality of delayed CSIT. As a result, we show that the DoF region with delayed imperfect-quality CSIT  is located between the DoF region with no CSIT and the DoF region with delayed CSIT.}

\maketitle

	\section{Introduction}
	
	Future cellular communication systems aim to provide ubiquitous and high-speed communication service for everywhere in the world, including railway, vehicles, submarines, drones, and etc. This unprecedented requirement brings many challenges to the communication system  design and analysis. One of these challenges is implementing the high-speed communication in the presence of high-mobility, since the fast time-varying wireless channel state, which is critical to high-speed communication, is very difficult to obtain at the transmitter side with timeliness guarantee. Therefore, it is reasonable to assume that the channel state information at the transmitter (CSIT) is delayed,  which refers to the CSIT reflect the past channel state information (CSI) but not the current CSI. Usually, delayed CSIT is possible, when the feedback speed is lagged behind the changing velocity of current CSI, e.g., in the high-mobility communication environment.
	
The research of estimating the average channel capacity in Rayleigh fading distribution can be found in \cite{0}. Aside from the statistical knowledge of channel, to analyze the fundamental performance limits and reveal useful insights of wireless networks  with delayed CSIT, the degrees-of-freedom (DoF) metric has been widely investigated in the past decade. In particular, DoF represents the multiplexing gain or capacity per-log factor, which tells us how many interference-free data streams can be decoded. In high signal-to-noise ratio (SNR) regime, one can believe that the achievable rate is proportional to DoF, and vice versa.  
	
	The research of DoF analysis with delayed CSIT stems from \cite{1}, where the DoF region was characterized for the $K$-user multiple-input single-output (MISO) broadcast channel with delayed CSIT, when  the number of transmit antennas, denoted by $M$, is equal to two for three-user scenario, and $K \le M$ for general $K$-user scenario. Furthermore, \cite{2} showed that the DoF region with delayed CSIT is located between the DoF region with no CSIT and the DoF region with perfect CSIT. Thereafter, the DoF region of two-user multiple-input multiple-output (MIMO) broadcast channel was derived in \cite{5}. For the three-user MIMO broadcast channel with delayed CSIT, the complete DoF region is unclear, where the latest progress on this problem can be found in \cite{3,4,5}. The DoF region of order-$(K-1)$ messages, which are desired by $K-1$ receivers, for the $K$-user MIMO broadcast channel with delayed CSIT was characterized in \cite{6}. 
	
	Reference \cite{7} first considered the DoF region of temporally-correlated MISO broadcast channel, where aside from delayed CSIT, the transmitter can estimate the current imperfect-quality CSI from delayed CSIT. Next, the DoF region of temporally-correlated MISO broadcast channel with delayed CSIT was completely characterized in \cite{8}. Not merely the imperfection come from the  current CSIT, the authors in \cite{9} considered the DoF region of MISO broadcast channel with imperfect-quality in both delayed and current  CSIT. However, the channel may not be temporally-correlated, thus the transmitter is blind of current CSI from delayed CSIT. To date, DoF with delayed imperfect-quality CSIT only is still unexplored.

	In this paper, we thus investigate the DoF region of two-user MIMO broadcast channel with delayed imperfect-quality CSIT, where the antenna configurations can be arbitrary. This CSIT model may be useful, when the feedback link has a delay and the CSI feedback has distortion.
	Specifically, we derive the \textit{optimal} DoF region by providing matched converse and achievability. The converse proof is based on the enhancement of physically degraded channel. The achievability proof is through a novel transmission scheme design. In contrast to \cite{1}-\cite{9}, the duration of each phase and the amount of transmitted symbols are configured based on the imperfection of delayed CSIT. The derived DoF region shows that the DoF region with delayed imperfect-quality CSIT lies between the DoF region with no CSIT and the DoF region with delayed CSIT. Moreover, the receiver having a less imperfection of CSIT enjoys a higher DoF value, for equal number of antennas at receivers.
	
	\textit{Notations}: Matrices and vectors are represented by upper and lowercase
	boldface letters, respectively.  $\log$ refers to $\log_2$. $\mathcal{O}(A)$ represents the same order of $A$. The block-diagonal matrix of blocks $\textbf{A},\textbf{B}$ is defined as $\mathcal{BD}\{\textbf{A},\textbf{B}\} \triangleq [\textbf{A},\textbf{0};\textbf{0},\textbf{B}]$.
	
	\section{System Model}

We consider a two-user $(M,N_1,N_2)$ MIMO broadcast channel, where transmitter $\text{Tx}$ has $M$ antennas and the receiver $\text{Rx}_i,i=1,2$ has $N_i$ antennas. 
The transmitter has two messages, i.e., $W_1,W_2$, where $W_1$ is desired by receiver $\text{Rx}_1$
and $W_2$ is desired by receiver $\text{Rx}_2$. At time slot $t$, the CSI matrix from transmitter $\text{Tx}$ to receiver $\text{Rx}_i$ is denoted by $\textbf{H}_i[t] \in \mathbb{C}^{N_i \times M}$, whose
elements are independent and identically distributed across
space and time, and drawn from a continuous distribution (e.g.,
Gaussian distribution for Rayleigh fading). The collection of CSI matrix from time slot $1$ to time slot $n$ is defined by $\mathcal{H}_i^n \triangleq \{\textbf{H}_i[1],\cdots,\textbf{H}_i[n]\}$. Mathematically, the input-output relationship at time slot $t$ can be expressed as 
\begin{equation}
	\textbf{y}_i[t] = \textbf{H}_i[t]\textbf{x}[t] + \textbf{n}_i,
\end{equation}
where the transmit signal at time slot $t$ is denoted by $\textbf{x}[t] \in \mathbb{C}^{M\times 1}$, the received signal at time slot $t$ is denoted by $\textbf{y}[t] \in \mathbb{C}^{N_i\times 1}$, and the additive white Gaussian
noise (AWGN) at receiver $\text{Rx}_i$ is denoted by $\textbf{n}_i \sim \mathcal{CN} (\textbf{0},\sigma^2\textbf{I}_{N_i})$. Moreover, the transmit signal is subject to an average maximal power constraint, i.e.,  $\frac{1}{n}\sum_{t=1}^n\textbf{x}^H[t]\textbf{x}[t] \le P$, where the maximal power is denoted by $P$.

The CSI feedback link is subject a delay, which is not less than one time slot. Moreover, the feedback link of CSI has distortion. Thus, at time slot $t$, only delayed imperfect-quality CSI, i.e., $\widehat{\mathcal{H}}_i^{t-1} \triangleq \{\widehat{\textbf{H}}_i[1],\cdots,\widehat{\textbf{H}}_i[t-1]\}$, is available at the transmitter. To be specific, CSI matrix can be written as 
\begin{equation}
	\textbf{H}_i[t] = 	\widehat{\textbf{H}}_i[t] + \widetilde{\textbf{H}}_i[t],
\end{equation}
where $\widehat{\textbf{H}}_i[t]$ and $\widetilde{\textbf{H}}_i[t]$ denote the imperfect-quality CSI and remaining part, respectively. $\widehat{\textbf{H}}_i[t]$ is with $ \mathcal{O}(\rho^{\alpha_i})$ power, where $\rho$ denotes the SNR and $\alpha_i \in [0,1]$ denotes the quality of CSIT. It can be seen that $\alpha_i = 1$ represents delayed CSIT, and $\alpha_i = 0$ represents no CSIT. As \cite{1}-\cite{9}, the CSI at receivers are perfect and global, namely each receiver has $\mathcal{H}_i^n,\forall i$.

The rate tuple is written as $(R_1(\rho),R_2(\rho))$, where rate $R_i = \frac{\log|\mathcal{W}_i|}{n}$ and $|\mathcal{W}_i|$ is the cardinality of message set $\mathcal{W}_i$. The encoding function $f(\cdot)$ encodes  $\textbf{x}[t] = f(W_1,W_2,\widehat{\mathcal{H}}_1^{t-1},\widehat{\mathcal{H}}_2^{t-1})$ at time slot $t$. The decoding function at receiver $\text{Rx}_i$, denoted by $g(\cdot)$, decodes $\widehat{W}_i = g(\textbf{y}[t],\mathcal{H}_1^{n},\mathcal{H}_2^{n})$ after $n$ time slots. The rate is said to be achievable, if there are a sequence of codebook pairs $\{\mathcal{B}_{1,t},\mathcal{B}_{1,t}\}_{t=1}^n$ and decoding functions $\{g_{1,n},g_{2,n}\}$ such that the error probabilities $P_{e}^{[n]}(\widehat{W}_i \ne W_i),\forall i$ go to zero when $n$ goes to infinity. The capacity region, denoted by $\mathcal{C}(\rho)$, is the region of all such achievable rate tuples. The DoF region is defined as the pre-log factor of the capacity region as $\rho \rightarrow {\cal{1}}$, 
\begin{equation} 
	\left\{ 
	(d_{1}, d_{2}) \in \mathbb{R}_+^2 \left| 
	\begin{split} 
		(R_{1}(\rho), R_{2}(\rho)) \in \mathcal{C}(\rho), \\
		d_{i} = \lim_{\rho \rightarrow {\cal{1}}} \frac{R_{i}(\rho)}{\log \rho}, i = 1,2.
	\end{split}\right.\right\}.
\end{equation}

\section{Main Results and Discussion} This section lists our mains and discussion.

\textbf{Theorem 1}: The DoF region of $(M,N_1,N_2)$ MIMO broadcast channel with delayed finite-precision CSIT, defined in Section-II, is given by
\begin{equation}
	\left\{
	\begin{aligned}
		(d_1,d_1) \\
		\in \mathbb{R}_+^2
	\end{aligned}
	\left|
	\begin{aligned}
		\dfrac{d_1}{\min\{N_1+\alpha_2N_2,M\}} 
		+ \dfrac{d_2}{\min\{N_2,M\}} \le 1 \\
		\dfrac{d_1}{\min\{N_1,M\}} + \dfrac{d_2}{\min\{N_2+\alpha_1N_1,M\}}   \le 1
	\end{aligned}
	\right\} \right. \nonumber
\end{equation}
\begin{proof}
	Please refer to Section-IV for the converse proof, and Section-VI for the achievability proof.
\end{proof}

\textbf{Remark 1}: Compared with the DoF region with no CSIT \cite{10} and the DoF region with delayed CSIT \cite{2}, one can see that the DoF region with delayed imperfect-quality CSIT lies between them. This is illustrated by Fig. 1, where it is on a $(2,1,1)$ MISO broadcast channel with $\alpha_1 = \alpha_2 = \alpha$.  Fig. 1 shows that increasing $\alpha$ enlarges the DoF region with delayed imperfect-quality CSIT.

\begin{figure}
	\centering
	\includegraphics[width=3in]{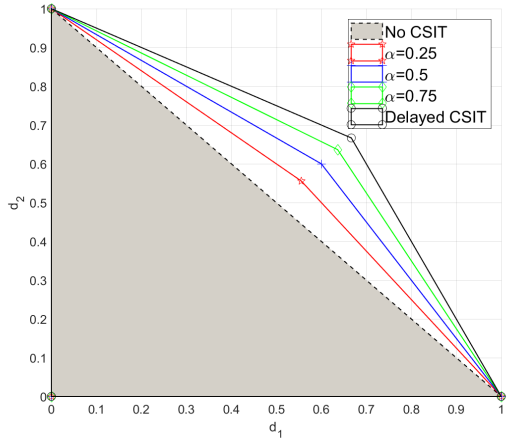}
	\caption{Comparison of DoF region: Proposed, No CSIT, Delayed CSIT.}
\end{figure}

\textbf{Proposition 1}: The corner points of DoF region in Theorem 1 is given as follows: $(0,0)$, $(0,\min\{N_2,M\})$, $(\min\{N_1,M\},0)$, and 
$
\left(\dfrac{ac(d-b)}{ad-bc}, \dfrac{bd(a - c)}{ad - bc} \right),
$ where $a = \min\{N_1,M\}$, $b = \min\{N_2 + \alpha_1N_1,M\}$, $c = \min\{N_1 + \alpha_2N_2,M\}$, and $d = \min\{N_2,M\}$.

\begin{proof}
	It can be easily seen that $(0,0)$, $(0,\min\{N_2,M\})$, and $(\min\{N_1,M\},0)$ are corner points on the coordinate. The strictly positive corner point $
	\left(\dfrac{ac(d-b)}{ad-bc}, \dfrac{bd(a - c)}{ad - bc} \right)
	$ can be proven via Matlab symbolic calculation for the linear system $d_1/a + d_2/b = 1,d_1/c + d_2/d = 1$.
\end{proof}

\textbf{Remark 2}: It can be seen that the strictly positive corner point $
\left(\dfrac{ac(d-b)}{ad-bc}, \dfrac{bd(a - c)}{ad - bc} \right) 
$ is complicated in expression. Thus, the achievability proof of  this corner point will involve tremendous calculations. To reduce the computation burden, we apply the transformation approach to directly obtain the achievable DoF region from the decoding condition. Furthermore, a higher quality of CSIT enjoys a better individual DoF. This is illustrated by Fig. 2, where it is on a $(2,1,1)$ MISO broadcast channel. Fig. 2 shows that $d_2$ beats $d_1$, if we gradually add the CSIT quality at receiver $\text{Rx}_2$.

\section{Proof of Theorem 1: The Converse}

The converse is proven via enhancing the original channel to a physically degraded channel and removing the delayed feedback constraint. 

To derive an outer region, there are two steps. In Step-I, a genie enhances the original channel by providing $\alpha_2N_2 \log \rho + \mathcal{O}(1)$ output equations at receiver $\text{Rx}_2$ to receiver $\text{Rx}_1$. Thus, the genie creates a physically degraded channel. 
In Step-II, as it is proven in \cite{11,12} that delayed feedback will not increase the capacity region with no CSIT, we can treat this physically degraded channel as a MIMO broadcast channel with no CSIT, where receiver $\text{Rx}_1$ has $N_1 + \alpha_2N_2$ antennas and receiver $\text{Rx}_2$ has $N_2$ antennas. According to the DoF region of MIMO broadcast channel with no CSIT, i.e., \cite[Theorem 1]{10}, the DoF region of original channel is outer-bounded as   
\begin{equation}
	\dfrac{d_1}{\min\{N_1+\alpha_2N_2,M\}} 
	+ \dfrac{d_2}{\min\{N_2,M\}} \le 1. \label{O1}
\end{equation}
Symmetrically, we have another outer region, which can be derived via similar two steps. 
In Step-I, a genie enhances the original channel by providing $\alpha_1N_1  \log \rho + \mathcal{O}(1)$ output equations at receiver $\text{Rx}_1$ to receiver $\text{Rx}_2$. Thus, the genie creates a physically degraded channel. 
In Step-II, as it is proven in \cite{11,12} that delayed feedback will not increase the capacity region with no CSIT, we can treat this physically degraded channel as a MIMO broadcast channel with no CSIT, where receiver $\text{Rx}_2$ has $N_2 + \alpha_1N_1$ antennas and receiver $\text{Rx}_1$ has $N_1$ antennas. According to the DoF region of MIMO broadcast channel with no CSIT, i.e., \cite[Theorem 1]{10}, the DoF region of original channel is outer-bounded as   
\begin{equation}
	\dfrac{d_1}{\min\{N_1,M\}} + \dfrac{d_2}{\min\{N_2+\alpha_1N_1,M\}}   \le 1.  \label{O2}
\end{equation}
The final outer region as desired is the union of \eqref{O1} and \eqref{O2}.

\section{An Illustrative Example for Achievability}

To begin with, we provide an illustrative example to highlight the novelty of proposed achievability, where the scenario is on the $(2,1,1)$ MISO broadcast channel with $\alpha_1 = \alpha_2 = \alpha$. Here, our aim is to prove the following achievable DoF region:
\begin{subequations}
	\begin{eqnarray}
		&&	d_1 + \frac{d_2}{1+\alpha} \le 1, \\
		&&	\frac{d_1}{1+\alpha} + d_2 \le 1.
	\end{eqnarray}
\end{subequations}
In fact, it suffices to show the corner point $(\frac{1+\alpha}{2+\alpha},\frac{1+\alpha}{2+\alpha})$ can be achieved by the proposed transmission scheme, given below.

\begin{figure}
	\centering
	\includegraphics[width=3in]{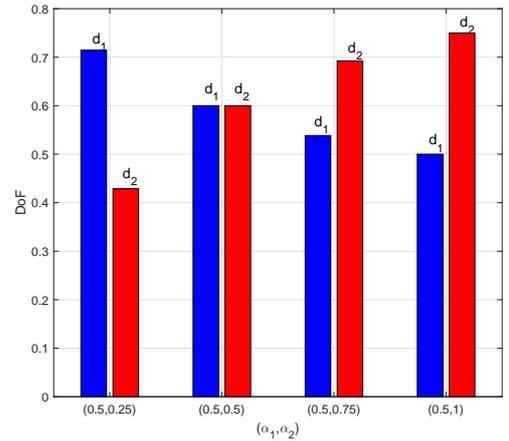}
	\caption{Comparison of different ($\alpha_1,\alpha_2$) pairs.}
\end{figure}

The proposed transmission scheme has three phases. In Phase-I, the  transmitter $\text{Tx}$ sends the symbols for receiver $\text{Rx}_1$, where the amount of symbols is larger the immediate decodability at receivers. In Phase-II, the transmitter $\text{Tx}$ sends the symbols for receiver $\text{Rx}_2$, where the amount of symbols is larger the immediate decodability at receivers. In Phase-III, the transmitter obtains delayed finite-precision CSIT. Thus, order-2 symbols, which are desired by two receivers, are construct at the transmitter and transmitted in this phase. It should be highlighted that in contrast to the existing design \cite{1}-\cite{9}, our novelty is to configure the duration of Phase-III, amount of symbols transmitted in Phase-I and II, according to CSIT quality. The details of the transmission scheme are elaborated below.

Phase-I spans $\tau$ time slots. During this phase,  transmitter $\text{Tx}$ sends $(1+\alpha)\tau$ symbols, i.e., $\textbf{s}_1 \in \mathbb{C}^{(1+\alpha)\tau \times 1}$, to receiver $\text{Rx}_1$, meanwhile transmitter $\text{Tx}$ does not send any symbols for receiver $\text{Rx}_2$. The received signals at each receiver can be given as follows:
\begin{subequations}
	\begin{eqnarray}
		&&	\textbf{y}_1^\text{P-I} = \textbf{H}_1^\text{P-I} \textbf{s}_1 + \textbf{n}_1, \\
		&&	\textbf{y}_2^\text{P-I} = \textbf{H}_2^\text{P-I} \textbf{s}_1 + \textbf{n}_2,
	\end{eqnarray}
\end{subequations}
where $\textbf{H}_i^\text{P-I} \triangleq \mathcal{BD}\{\textbf{h}_i[1],\cdots,\textbf{h}_i[\tau]\},i=1,2$.

Phase-II spans $\tau$ time slots. During this phase,  transmitter $\text{Tx}$ sends $(1+\alpha)\tau$ symbols, i.e., $\textbf{s}_2 \in \mathbb{C}^{(1+\alpha)\tau \times 1}$, to receiver $\text{Rx}_2$, meanwhile transmitter $\text{Tx}$ does not send any symbols for receiver $\text{Rx}_1$. The received signals at each receiver can be given as follows:
\begin{subequations}
	\begin{eqnarray}
		&&	\textbf{y}_1^\text{P-II} = \textbf{H}_1^\text{P-II} \textbf{s}_2 + \textbf{n}_1, \\
		&&	\textbf{y}_2^\text{P-II} = \textbf{H}_2^\text{P-II} \textbf{s}_2 + \textbf{n}_2,
	\end{eqnarray}
\end{subequations}
where $\textbf{H}_i^\text{P-II} \triangleq \mathcal{BD}\{\textbf{h}_i[\tau +1],\cdots,\textbf{h}_i[2\tau]\}, i=1,2$.

Phase-III spans $\alpha\tau$ time slots. Now, the transmitter obtains the incomplete CSI of Phase-I and Phase-II, i.e., $\widehat{\textbf{H}}_i^\text{P-I}$ and $\widehat{\textbf{H}}_i^\text{P-II}$. Thereby, the transmitter can construct the following order-2 symbols:
\begin{eqnarray}
	&& {\bm{\eta}} \triangleq  \widehat{\textbf{H}}_2^\text{P-I}\textbf{s}_1 + \widehat{\textbf{H}}_1^\text{P-II}\textbf{s}_2.
\end{eqnarray}
Note that if $\textbf{s}_1$ and $\textbf{s}_2$ are from lattice, then ${\bm{\eta}}$ will be a valid lattice codeword. Each $\widehat{\textbf{H}}_2^\text{P-I}$ and $\widehat{\textbf{H}}_1^\text{P-II}$ contains $\alpha \tau \log \rho + \mathcal{O}(1)$ equations, due to  $\widehat{\textbf{H}}_2^\text{P-I}, \widehat{\textbf{H}}_1^\text{P-II} \sim \mathcal{O}(\rho^{\alpha})$. One can send ${\bm{\eta}}$ using $\alpha \tau$ time slots, where one transmit antenna is used at each time. 

Thus, it can be seen that the achievable DoF for each receiver is $(1+\alpha)\tau / (2\tau + \alpha\tau) = (1+\alpha) / (2 + \alpha)$.

\section{Proof of Theorem 1: The Achievability} 
The achievability is proven from two cases.
\textit{$M \le N_2$ Case}:
In this case, it suffices to prove the following achievable DoF region, given by
\begin{subequations}
	\begin{eqnarray}
		\dfrac{d_1}{M} + \frac{d_2}{M} \le 1,\\
		\frac{d_1}{\min\{N_1,M\}} + \dfrac{d_2}{M} \le 1.
	\end{eqnarray}
\end{subequations}

This region can be achieved by time-sharing of the following time division multiple-access (TDMA) schemes: The first scheme is that the transmitter $\text{Tx}$ sends $\min\{N_1,M\}$ symbols for receiver $\text{Rx}_1$ in one time slot, where the  receiver $\text{Rx}_1$ can immediate decode the transmitted symbols. The second scheme is that the transmitter $\text{Tx}$ sends $M$ symbols for receiver $\text{Rx}_2$ in one time slot, where the  receiver $\text{Rx}_2$ can immediate decode the transmitted symbols. It can be seen that no CSIT is required for the above schemes.

\textit{$N_2 < M$ Case}:
In this much general case, there are a few challenges ahead, listed as follows: 1) The asymmetry of antenna configurations and CSIT quality; and 2) the non-trivial analysis of achievable DoF region of the scheme.
To generalize the idea in the illustrative example, we use the following techniques to overcome the challenges: 1) Parameterized configuration of each phase duration, amount of symbols transmitted in Phase-I and II; and 2) Deriving the achievable DoF region by transforming the decoding condition.
Next, the details of the transmission scheme are elaborated below.

Phase-I spans $\tau_1$ time slots. During this phase, the transmitter $\text{Tx}$ sends $\min\{N_1+\alpha_2N_2,M\}\tau_1$ symbols, i.e., $\textbf{s}_1 \in \mathbb{C}^{\min\{N_1+\alpha_2N_2,M\}\tau_1 \times 1}$, to receiver $\text{Rx}_1$, meanwhile transmitter $\text{Tx}$ does not send any symbols for receiver $\text{Rx}_2$. The received signals at each receiver can be given as follows:
\begin{subequations}
	\begin{eqnarray}
		&&	\textbf{y}_1^\text{P-I} = \textbf{H}_1^\text{P-I} \textbf{s}_1 + \textbf{n}_1, \\
		&&	\textbf{y}_2^\text{P-I} = \textbf{H}_2^\text{P-I} \textbf{s}_1 + \textbf{n}_2,
	\end{eqnarray}
\end{subequations}
where $\textbf{H}_i^\text{P-I} \triangleq \mathcal{BD}\{\textbf{h}_i[1],\cdots,\textbf{h}_i[\tau_1]\},i=1,2$.

Phase-II spans $\tau_2$ time slots. During this phase,  transmitter $\text{Tx}$ sends $\min\{N_2+\alpha_1N_1,M\}\tau_2$ symbols, i.e., $\textbf{s}_2 \in \mathbb{C}^{\min\{N_2+\alpha_1N_1,M\}\tau_2 \times 1}$, to receiver $\text{Rx}_2$, meanwhile transmitter $\text{Tx}$ does not send any symbols for receiver $\text{Rx}_1$. The received signals at each receiver can be given as follows:
\begin{subequations}
	\begin{eqnarray}
		&&	\textbf{y}_1^\text{P-II} = \textbf{H}_1^\text{P-II} \textbf{s}_2 + \textbf{n}_1, \\
		&&	\textbf{y}_2^\text{P-II} = \textbf{H}_2^\text{P-II} \textbf{s}_2 + \textbf{n}_2,
	\end{eqnarray}
\end{subequations}
where $\textbf{H}_i^\text{P-II} \triangleq \mathcal{BD}\{\textbf{h}_i[\tau_1 +1],\cdots,\textbf{h}_i[\tau_1+\tau_2]\}, i=1,2$.

Phase-III spans $\tau_3$ time slots. So far, the transmitter obtains the incomplete CSI of Phase-I and Phase-II, i.e., $\widehat{\textbf{H}}_i^\text{P-I}$ and $\widehat{\textbf{H}}_i^\text{P-II}$. Thereby, the transmitter can construct the following order-2 symbols:
\begin{eqnarray}
	&& {\bm{\eta}} \triangleq  \widehat{\textbf{H}}_2^\text{P-I}\textbf{s}_1 + \widehat{\textbf{H}}_1^\text{P-II}\textbf{s}_2.
\end{eqnarray}
Note that if $\textbf{s}_1$ and $\textbf{s}_2$ are from lattice, then ${\bm{\eta}}$ will be a valid lattice codeword. $\widehat{\textbf{H}}_2^\text{P-I}$   contains $\alpha_2 N_2 \tau_1 \log \rho + \mathcal{O}(1)$ equations, due to  $\widehat{\textbf{H}}_2^\text{P-I} \sim \mathcal{O}(\rho^{\alpha_2})$.
$\widehat{\textbf{H}}_1^\text{P-II}$   contains $\alpha_1 N_1 \tau_2 \log \rho + \mathcal{O}(1)$ equations, due to  $\widehat{\textbf{H}}_1^\text{P-II} \sim \mathcal{O}(\rho^{\alpha_1})$. The receiver $\text{Rx}_1$ can decode $N_1\tau_3 \log \rho + \mathcal{O}(1)$ equations within Phase-III without any delay. Since receiver $\text{Rx}_1$ needs $(\min\{N_1+\alpha_2N_2,M\}\tau_1 -   N_1\tau_1) \log \rho + \mathcal{O}(1)$ equations, we have to follow that
\begin{equation}
	\min\{N_1+\alpha_2N_2,M\}\tau_1 -   N_1\tau_1 \le N_1\tau_3. \label{E1}
\end{equation}
The receiver $\text{Rx}_2$ can decode $N_2\tau_3 \log \rho + \mathcal{O}(1)$ equations within Phase-III without any delay. Since receiver $\text{Rx}_2$ needs $(\min\{N_2+\alpha_1N_1,M\} \tau_2 - N_2\tau_2)\log \rho + \mathcal{O}(1)$ equations, we have to follow that
\begin{equation}
	\min\{N_2+\alpha_1N_1,M\}\tau_2 - N_2\tau_2 \le N_2\tau_3.\label{E2}
\end{equation}
As inequalities \eqref{E1} and \eqref{E2} depend on the decodability of proposed transmission scheme, we name them by decoding condition of the scheme.

We will transform the achievable DoF region from the decoding condition, i.e., \eqref{E1} and \eqref{E2}. First, we add $N_1\tau_1 + N_1\tau_2$ at both sides of \eqref{E1} and $N_2\tau_1 + N_2\tau_2$ at both sides of \eqref{E2}. This yields,
\begin{subequations}
	\begin{eqnarray}
		\min\{N_1 + \alpha_2N_2,M\}\tau_1 + N_1\tau_2 \le N_1(\tau_1+\tau_2+\tau_3), \label{E3}\\
		\min\{N_2+\alpha_1N_1,M\}\tau_2 + N_2\tau_1 \le N_2(\tau_1+\tau_2+\tau_3).  \label{E4}
	\end{eqnarray}
\end{subequations}
Next, we divide both sides of \eqref{E3} by $N_1(\tau_1+\tau_2+\tau_3)$ and both sides of \eqref{E4} by $N_2(\tau_1+\tau_2+\tau_3)$, i.e.,
\begin{subequations}
	\begin{eqnarray}
		\dfrac{\min\{N_1 + \alpha_2N_2,M\}\tau_1}{N_1(\tau_1+\tau_2+\tau_3)} + \dfrac{\tau_2}{\tau_1+\tau_2+\tau_3} \le 1, \label{E5}\\
		\dfrac{\min\{N_2+\alpha_1N_1,M\}\tau_2}{N_2(\tau_1+\tau_2+\tau_3)} + \dfrac{\tau_1}{\tau_1+\tau_2+\tau_3} \le 1.  \label{E6}
	\end{eqnarray}
\end{subequations}
Since the achievable DoF tuple $(d_1,d_2)$ can be expressed as 
\begin{equation}
	\left(\dfrac{\min\{N_1 + \alpha_2N_2,M\}\tau_1}{\tau_1+\tau_2+\tau_3}, \dfrac{\min\{N_2 + \alpha_1N_1,M\}\tau_2}{\tau_1+\tau_2+\tau_3}\right),
	\nonumber
\end{equation}
we can transform \eqref{E5} and \eqref{E6} into the achievable DoF region, given by
\begin{subequations}
	\begin{eqnarray}
		\dfrac{d_1}{\min\{N_1+\alpha_2N_2,M\}} 
		+ \dfrac{d_2}{N_2} \le 1, \\
		\dfrac{d_1}{N_1} + \dfrac{d_2}{\min\{N_2+\alpha_1N_1,M\}}   \le 1.
	\end{eqnarray}
\end{subequations}

\section{Conclusion}

In this paper, we characterized the DoF region of two-user $(M,N_1,N_2)$ MIMO broadcast channel with delayed imperfect-quality CSIT. The matched converse and achievability proofs were provided, where a novel transmission scheme was designed and analyzed by transformation. Through this study, we  showed that  the DoF region with delayed imperfect-quality CSIT  is located between the DoF region with no CSIT and the DoF region with delayed CSIT, and a higher quality of CSIT enjoys a better individual DoF, for equal number of antennas at receivers. In the future, it is interesting to extend this work to the three-user scenario.

\vskip5pt

\noindent  \textit{Funding Information}: This work was supported by the China Postdoctoral Science Foundation under grant no.
2021M691453.

\vskip5pt

\noindent \textit{Conflict of interest}: No financial interests/personal relationships need to be considered as potential competing interests.

\vskip5pt

\noindent  \textit{Data availability statement}: Data sharing not applicable to this articles no datasets were generated or analysed during the current study.

\vskip5pt

\noindent T. Zhang (\textit{Southern University of Science and Technology, China})
\vskip3pt

\noindent E-mail: zhangt7@sustech.edu.cn

\end{document}